\documentclass{article}

\usepackage{multicol}
\usepackage{comment}
\usepackage{amsmath}
\usepackage{enumitem}
\usepackage{url}
\usepackage{syntax}
\usepackage[dvipdfmx]{graphicx}
\usepackage{authblk}

\begin{document}

\title{Non-Linear Pattern-Matching against Unfree Data Types with Lexical Scoping}

\author{Satoshi Egi}

\maketitle

\begin{abstract}
This paper proposes a pattern-matching system that enables non-linear pattern-matching against unfree data types.
The system allows multiple occurrences of the same variables in a pattern, multiple results of pattern-matching and modularization of the way of pattern-matching for each data type at the same time.
It enables us to represent pattern-matching against not only algebraic data types but also \textit{unfree data types} such as sets, graphs and any other data types whose data have no canonical form and multiple ways of decomposition.
I have realized that with a rule that pattern-matching is executed from the left side of a pattern and a rule that a binding to a variable in a pattern can be referred to in its right side of the pattern.
Furthermore, I have realized modularization of these patterns with lexical scoping.
In my system, a pattern is not a first class object, but a \textit{pattern-function} that obtains only patterns and returns a pattern is a first class object.
This restriction simplifies the non-linear pattern-matching system with lexical scoping.
I have already implemented the pattern-matching system in the Egison programming language.
\end{abstract}

\section{Introduction}

In this paper, I focus on the representation of non-linear pattern-matching against unfree data types.
Unfree data types are data types whose data have no \textit{canonical form}.
A canonical form of an object is a standard way to represent that object.
For instance, data of sets and graphs do not have a canonical form.
For example, a collection \texttt{\{a, b, c\}} is equal to \texttt{\{b, a, c\}}, \texttt{\{c, b, a\}} and \texttt{\{a, a, b, c\}}, if it is regarded as a set.

Data types whose data have no canonical form often play important roles in expressing algorithms.
Then, a natural way to handle these kinds of data is really important.
Without it, we need to translate and regard them as a data type whose data have a canonical form when we treat them.
For example, a set would be treated as a list.
Many programmers think this is unavoidable, and in fact, it is a latent stress of programming.

I have designed and implemented a new pattern-matching system that allows multiple occurrences of the same variables in a pattern, multiple results of pattern-matching and modularization of the way of pattern-matching for each data type at the same time.
With this system, we can treat unfree data types directly.
I will show its expressive power and implementation in this paper.

Let me demonstrate my pattern-matching system.
I use Haskell-like pseudocode in this section.
First, I introduce the \texttt{matchAll} expression to explain the concept of my pattern-matching system.
We can understand the concept with only this expression.

{\footnotesize
\begin{verbatim}
matchAll
  [1, 2, 3] as        -- Target  (`as' is a reserved word)
  (List Integer) with -- Matcher (`with' is a reserved word)
  x:xs -> (x, xs)     -- Match-Clause: Pattern -> Body
--> [(1,[2,3])]       -- Result
\end{verbatim}
}

The characteristic of this expression is it takes a \textit{matcher}.
The interpreter matches the target with the pattern in the way specified with the matcher.
A matcher specifies the way of pattern-matching.
We can modularize the way of pattern-matching for each data type in matchers.

In the following example, we pattern-match against a collection as a list, multiset, set respectively.
The pattern constructor `\texttt{:}' (cons) divides a collection into an element and the rest.
The meaning of `\texttt{:}' varies for each matcher and is expressed in the matcher definition.
I explain how we define matchers in section~\ref{matcher}.

{\footnotesize
\begin{verbatim}
matchAll [1, 2, 3] as (List Integer) with x:xs -> (x, xs)
--> [(1,[2,3])]
matchAll [1, 2, 3] as (Multiset Integer) with x:xs -> (x, xs)
--> [(1,[2,3]),(2,[1,3]),(3,[1,2])]
matchAll [1, 2, 3] as (Set Integer) with x:xs -> (x, xs)
--> [(1,[1,2,3]),(2,[1,2,3]),(3,[1,2,3])]
\end{verbatim}
}

Note that we are handling pattern-matching with multiple results.
This feature is necessary to pattern-match against data types whose data have no standard form such as multisets and sets.
This is because they have multiple ways of decomposition.
We need backtracking in the pattern-matching process to try multiple forms.

The pattern constructor `\texttt{++}' (join) divides a collection into two collections.
Lists also have the multiple ways of decomposition if we introduce this pattern constructor.

{\footnotesize
\begin{verbatim}
matchAll [1, 2, 3] as (List Integer) with xs ++ ys -> (xs, ys)
--> [([],[1,2,3]),([1],[2,3]),([1,2],[3]),([1,2,3],[])]
\end{verbatim}
}

Non-linear pattern-matching is also important to represent pattern-matching against unfree data types.
Non-linear pattern-matching with backtracking eliminates deeply nested loops and conditional branches.
For example, non-linear patterns enable us to represent a pattern that matches when the collection has multiple same elements.

{\footnotesize
\begin{verbatim}
matchAll [1, 2, 3, 3, 2, 4] as (Multiset Integer) with x:x:_ -> x
--> [2,3,3,2]
\end{verbatim}
}

A part of the pattern that does not contain pattern constructors and free variables is a \textit{value-pattern}.
In most of cases, the equality is checked for a value-pattern.
We can define the equality for each data type in the matcher.
In the above sample, the second appearance of `\texttt{x}' is a value-pattern.

My system can handle pattern-matching against infinite collections with infinite results.
The following sample enumerates all twin primes from the infinite list of prime numbers.
In the following sample, the second appearance of `\texttt{p + 2}' is a value pattern.
Note that we can write an any expression in a value-pattern.

{\footnotesize
\begin{verbatim}
take 8 (matchAll primes as (List Integer) with
          (_ ++ (p:p + 2:_))) -> (p, p + 2))
--> [(3,5),(5,7),(11,13),(17,19),(29,31),(41,43),(59,61),(71,73)]
\end{verbatim}
}

We can pattern-match also against nested data types such as lists of sets and sets and sets.
The following code enumerates the common elements among the collections inside a collection.

{\footnotesize
\begin{verbatim}
matchAll [[1, 2, 3, 4, 5], [4, 5, 1] [6, 1, 7, 4]]
      as (List (Multiset Integer)) with
  (n:_):(n:_):(n:_):[] -> n
--> [1,4]
\end{verbatim}
}

Non-linear pattern-matching with backtracking is really challenging.
I had to invent new syntax and a new mechanism for that.
I have realized that with a rule that pattern-matching is executed from the left side of a pattern and a rule that a binding to a variable in a pattern can be referred to in its right side of the pattern.

Furthermore, I have realized lexical scoping in patterns to modularize useful patterns in many places of programs.
Lexical scoping in patterns becomes a challenging problem because my system allows non-linear patterns.

I have solved all problems and created the new programming language Egison.
In this paper, I introduce my new pattern-matching system by showing various programs in my new language and its implementation.
I have implemented it using Haskell.

\section{Preliminaries}

Before explaining pattern-matching, I introduce basics of my language to understand the rest of this paper.
Actually, samples in the previous section are just pseudocode and my language has completely different syntax.
My language is a purely functional programming language with a strong pattern-matching facility.
In this section, I explain the ordinary purely functional aspect of the language.
I explain patterns and pattern-matching from the next section.

I demonstrate pattern-matching on the interpreter as below.
`\texttt{>}' is a prompt.
An expression after a prompt is input.
Output is displayed from the next line of the end of input.
My language has parenthesized syntax as Lisp.
We can add top-level bindings from the prompt with a \texttt{define} expression.
Bindings added by a \texttt{define} expression can be referred from the next prompt.

{\footnotesize
\begin{verbatim}
> (define $x 10)
> x
10
> (+ x 100)
110
\end{verbatim}
}

\subsection{Built-in Data}

In this paper, I use only booleans and integers for built-in data.
Booleans are represented as `\texttt{\#t}' and `\texttt{\#f}'.
We can represent numbers as other programming languages.
For example, my language represent negative numbers by adding `\texttt{-}' ahead of a number literal as `\texttt{-123}'.

\subsection{Objects}

\subsubsection{Inductive Data}

We can construct a complex object using \textit{inductive data}.
An inductive datum consists of a data constructor and its arguments enclosed with angled-brackets.
It can have any inductive data as arguments.
This is why it is called inductive.
Note that the name of a constructor has to start with uppercase.

{\footnotesize
\begin{verbatim}
> <Nil>
<Nil>
> <Cons 1 <Cons 2 <Nil>>>
<Cons 1 <Cons 2 <Nil>>>
\end{verbatim}
}

\subsubsection{Tuples (Multiple Values)}

A tuple is expressed as a sequence of elements enclosed in square brackets.
Note that a tuple consists of a single element is treated as the same object with the element itself.

{\footnotesize
\begin{verbatim}
> [1 [[2]]]
[1 2]
\end{verbatim}
}

\subsubsection{Collections}

A collection is a sequence of elements enclosed in braces.
Note that an expression that has `\texttt{@}' is dealt not as an element but a \textit{subcollection}, a segment of the collection.
Using this notation, we can combine collections.

{\footnotesize
\begin{verbatim}
> {@{@{1}} @{2 @{3}} 4}
{1 2 3 4}
\end{verbatim}
}

\subsubsection{Functions}

We define a function using \texttt{lambda} as other functional programming languages.

\subsubsection{Pattern-Functions}

A pattern-function is a function that gets only patterns and returns a pattern.
We define it using a \texttt{pattern-function} expression.
We demonstrate a lot of patterns and pattern-functions from the next section.

\subsubsection{Matchers}

A matcher is defined to specify how to pattern-match for each data type.
We define it using a \texttt{matcher} expression.
We explain these expressions in detail from the next section.

\subsection{Syntax}

My language has \texttt{if}, \texttt{let}, and \texttt{letrec} expressions as other functional programming languages.
I omit explanation about these expressions.

\texttt{match-all} and \texttt{match} expressions are syntax for pattern-matching, the core of this paper.
I explain these expressions in detail from the next section.

\section{Pattern-Matching Expressions}\label{exprs}

This section explains how we express pattern-matching and demonstrates its expressive power.

\subsection{Pattern-Matching with Backtracking}

The following is syntax of the \texttt{match-all} expression.
A \texttt{match-all} expression is composed of a \textit{target}, \textit{matcher} and \textit{match-clause}, which consists of a \textit{pattern} and \textit{body expression}.
A \texttt{match-all} expression evaluates the body of the match-clause for each pattern-matching result and returns the collections that contains all results.
A matcher specifies the way to match the target with the pattern.

\begin{grammar}
<match-all-expr> ::= `(match-all' <tgt-expr> <matcher-expr> <match-clause> `)'

<match-clause> ::= `[' <pattern> <expr> `]'
\end{grammar}

The following is the first samples of the \texttt{match-all} expression.
The only difference among these three expressions is its matcher.

{\footnotesize
\begin{verbatim}
> (match-all {1 2 3} (list integer) [<cons $x $ts> [x ts]])
{[1 {2 3}]}
> (match-all {1 2 3} (multiset integer) [<cons $x $ts> [x ts]])
{[1 {2 3}] [2 {1 3}] [3 {1 2}]}
> (match-all {1 2 3} (set integer) [<cons $x $ts> [x ts]])
{[1 {1 2 3}] [2 {1 2 3}] [3 {1 2 3}]}
\end{verbatim}
}

\texttt{<cons \$x \$ts>} is a \textit{inductive-pattern}.
\texttt{cons} is a pattern-constructor.
The name of a pattern-constructor starts with lowercase.
The pattern-constructor \texttt{cons} takes patterns as arguments.
It divides a collection into a head element and the rest.
The meaning of a head differs for each matcher.
For example, multisets ignore the order of the elements of a collection, so every element can be the head element.
`\texttt{\$x}' and `\texttt{\$ts}' are called \textit{pattern-variables}.
We can access the result of pattern-matching referring to them.

Sets ignore the order and the duplicates of the elements of a collection.
Therefore, the target collection itself is bounded to `\texttt{ts}'.

We can deal with pattern-matching that has infinite results.
I explain this mechanism in section~\ref{mechanism} in detail.

{\footnotesize
\begin{verbatim}
> (take 10 (match-all nats (set integer) [<cons $m <cons $n _>> [m n]]))
{[1 1] [1 2] [2 1] [1 3] [2 2] [3 1] [1 4] [2 3] [3 2] [4 1]}
\end{verbatim}
}

\texttt{take} is a function that gets a number \textit{n} and a collection \textit{xs} and returns the first \textit{n} elements of \textit{xs}.
\texttt{nats} is an infinite list that contains all natural numbers.
`\texttt{_}' is an wildcard and matches with any object.
Note that we extract two elements from the collection with the nested \texttt{cons} inductive pattern.

I introduces other pattern-constructors \texttt{nil} and \texttt{join}.
The \texttt{nil} pattern-constructor takes no arguments and matches when the target is an empty collection.
The \texttt{join} pattern-constructor takes two arguments and divides a collection into two collections.

The following is a demonstration of \texttt{join}.

{\footnotesize
\begin{verbatim}
> (match-all {1 2 3} (list integer) [<join $xs $ys> [xs ys]])
{[{} {1 2 3}] [{1} {2 3}] [{1 2} {3}] [{1 2 3} {}]}
\end{verbatim}
}

The order of the elements in the result of pattern-matching is very important.
Specifically, why is not the result of the first expression as follow?

{\footnotesize
\begin{verbatim}
{[{1 2 3} {}] [{1 2} {3}] [{1} {2 3}] [{} {1 2 3}]}
\end{verbatim}
}

The reason is there is a case that the target collection is an infinite list.
If we adopted the latter order, the interpreter tries to find the last element from the infinite list of natural numbers and cannot return any answers for the following pattern-matching.

{\footnotesize
\begin{verbatim}
> (take 10 (match-all nats (list integer) [<join _ <cons $x _>> x]))
{1 2 3 4 5 6 7 8 9 10}
\end{verbatim}
}

\subsection{Non-Linear Pattern-Matching}

Non-linear pattern-matching is one of the most important features of my pattern-matching system.
Non-linear pattern-matching is necessary to represent meaningful patterns against unfree data types.
The following is an example of a non-linear pattern.
The output of this example is the collection of numbers from which three number sequence starts.

{\footnotesize
\begin{verbatim}
> (match-all {1 5 6 2 4} (multiset integer)
    [<cons $n <cons ,(+ n 1) <cons ,(+ n 2) _>>> n])
{4}
\end{verbatim}
}

Pattern-matching is executed from left to right, and the binding to a pattern-variable can be referred to in its right side of the pattern.
In this example, at first, the pattern-variable `\texttt{\$n}' is bound to any element of the collection.
After that, the value-pattern `\texttt{,(+ n 1)}' and `\texttt{,(+ n 2)}' are examined.
A value-pattern has `\texttt{,}' ahead of it.
The expression following `\texttt{,}' can be any kind of expressions.
A value-pattern is a pattern that matches if the object is equal with the content of the pattern.
The meaning of ``equal'' is defined in matchers, and then varies by matchers.
`\texttt{,(+ n 1)}' and `\texttt{,(+ n 2)}' place the right side of `\texttt{\$n}'.
Therefore, after successful pattern-matching, `\texttt{\$n}' is bound to an element from which three number sequence starts.

A value-pattern is one of the most important inventions of my proposal.
Guard notation is not good with my system.
This is because guard conditions are checked after pattern-matching.
Therefore, it causes unnecessary backtracking in the middle of the pattern-matching process.

The following code is the second example of non-linear pattern-matching.
It enumerates all twin primes from the infinite list of prime numbers with pattern-matching.

{\footnotesize
\begin{verbatim}
> (define $twin-primes
    (match-all primes (list integer)
      [<join _ <cons $p <cons ,(+ p 2) _>>> [p (+ p 2)]]))
> (take 8 twin-primes)
{[3 5] [5 7] [11 13] [17 19] [29 31] [41 43] [59 61] [71 73]}
\end{verbatim}
}

We can write pattern-matching against nested unfree data types such as a list of multisets or a set of sets as the following sample code.

{\footnotesize
\begin{verbatim}
> (match-all {{1 2 3 4 5} {4 5 1} {6 1 7 4}} (list (multiset integer))
    [<cons <cons $n _> <cons <cons ,n _> <cons <cons ,n _> <nil>>>> n])
{1 4}
\end{verbatim}
}

My language has also the \texttt{match} expression as other functional languages.
A \texttt{match} expression takes multiple match-clauses and tries pattern-matching for each pattern from the head of match-clauses.
A \texttt{match} expression is useful to express conditional branches.

\begin{grammar}
<match-expr> ::= `(match' <tgt-expr> <matcher-expr> `\{' <match-clause>* `\})' 
\end{grammar}

Figure~\ref{fig:poker} is a demonstration program that determines poker-hands.
Note that all poker-hands are represented in a single pattern.
I explain the definition of the \texttt{mod}, \texttt{suit} and \texttt{card} matcher in section~\ref{matcher}.

\begin{figure}
  \begin{center}

    {\scriptsize
\begin{verbatim}
\end{verbatim}
  }

\begin{multicols}{2}
  
  {\scriptsize
\begin{verbatim}
;;
;; Matcher definitions
;;
(define $mod
 (lambda [$m]
  (matcher
    {[,$n []
      {[$tgt (if (eq? (modulo tgt m) (modulo n m))
                 {[]}
                 {})]}]
     [$ [something] {[$tgt {tgt}]}]
     })))

(define $suit
  (algebraic-data-matcher
    {<spade> <heart> <club> <diamond>}))

(define $card
  (algebraic-data-matcher
    {<card suit (mod 13)>}))

;;
;; A function that determines poker-hands
;;
(define $poker-hands
  (lambda [$cs]
    (match cs (multiset card)
      {[<cons <card $s $n>
         <cons <card ,s ,(- n 1)>
          <cons <card ,s ,(- n 2)>
           <cons <card ,s ,(- n 3)>
            <cons <card ,s ,(- n 4)>
             <nil>>>>>>
        <Straight-Flush>]
       [<cons <card _ $n>
         <cons <card _ ,n>
          <cons <card _ ,n>
            <cons <card _ ,n>
              <cons _
                <nil>>>>>>
        <Four-of-Kind>]
       [<cons <card _ $m>
         <cons <card _ ,m>
          <cons <card _ ,m>
           <cons <card _ $n>
            <cons <card _ ,n>
              <nil>>>>>>
        <Full-House>]
       [<cons <card $s _>
         <cons <card ,s _>
           <cons <card ,s _>
             <cons <card ,s _>
               <cons <card ,s _>
                 <nil>>>>>>
        <Flush>]
\end{verbatim}
  }

  \columnbreak

  {\scriptsize
\begin{verbatim}
       [<cons <card _ $n>
         <cons <card _ ,(- n 1)>
          <cons <card _ ,(- n 2)>
           <cons <card _ ,(- n 3)>
            <cons <card _ ,(- n 4)>
             <nil>>>>>>
        <Straight>]
       [<cons <card _ $n>
         <cons <card _ ,n>
          <cons <card _ ,n>
           <cons _
            <cons _
             <nil>>>>>>
        <Three-of-Kind>]
       [<cons <card _ $m>
         <cons <card _ ,m>
          <cons <card _ $n>
            <cons <card _ ,n>
             <cons _
               <nil>>>>>>
        <Two-Pair>]
       [<cons <card _ $n>
         <cons <card _ ,n>
          <cons _
           <cons _
            <cons _
             <nil>>>>>>
        <One-Pair>]
       [<cons _
         <cons _
          <cons _
           <cons _
            <cons _
             <nil>>>>>>
        <Nothing>]})))
\end{verbatim}
  }

\end{multicols}

  \end{center}
  \caption{Pattern-matching to determine poker hands}
  \label{fig:poker}
\end{figure}

\subsection{Or-Patterns, And-Patterns, Not-Patterns and Tuple-Patterns}

We can enumerate prime triplets with pattern-matching using an \textit{or-pattern} and \textit{and-pattern}.
An or-pattern matches with the object, if the object matches one of given patterns.
An and-pattern is a pattern that matches the object, if and only if all of the patterns are matched.
An and-pattern is used like an \textit{as-pattern} in the following code that enumerates prime triplets.

{\footnotesize
\begin{verbatim}
> (define $prime-triplets
    (match-all primes (list integer)
      [<join _ <cons $p
                <cons (& $m (| ,(+ p 2) ,(+ p 4)))
                 <cons ,(+ p 6) _>>>>
       [p m (+ p 6)]]))
> (take 7 prime-triplets)
{[5 7 11] [7 11 13] [11 13 17] [13 17 19] [17 19 23] [37 41 43] [41 43 47]}
\end{verbatim}
}

A \textit{not-pattern} matches with a object if the object does not matches the pattern following after `\texttt{\^}'.
The following sample enumerates the elements of the collection that appears only once.

{\footnotesize
\begin{verbatim}
> (match-all {1 2 3 3 2 4} (multiset integer) [<cons $x ^<cons ,x _>> x])
{1 4}
\end{verbatim}
}

We can write a tuple in a pattern.
Such a pattern is called a \textit{tuple-pattern}.
When a pattern is a tuple-pattern, the target and matcher must be also tuples.
The following is a sample for a tuple-pattern.

{\footnotesize
\begin{verbatim}
> (match-all [3 3] [integer integer]
   [[$n ,n] n])
{3}
\end{verbatim}
}

\subsection{Modularization of Patterns with Lexical Scoping}

Modularization of patterns is a necessary feature to reuse useful patterns.
Non-linear patterns make modularization of patterns difficult.

Patterns are not first class objects in my pattern-matching system.
Therefore, for example, `\texttt{(define \$x \_)}' is illegal, because `\texttt{\_}' is a pattern and not a first class object.
However, a \textit{pattern-function}, a function that takes patterns and returns a pattern, is a first class object.
We can define pattern-functions in anywhere of programs, and use them to generate patterns or define other patter-functions.

\begin{grammar}
<pat-func-expr> ::= `(pattern-function [' <pat-var>* `]' <pattern'> `)'
\end{grammar}

Since a pattern-function has lexical scoping as a normal function by \texttt{lambda}, the bindings for the pattern-variables in the argument patterns and the body of pattern-functions don't conflict.
Then, we don't have to care about which pattern-variables occur in a pattern-function.
In the following sample, what is bound to `\texttt{\$m}' and `\texttt{\$n}' don't matter in the body of the pattern-function \texttt{twin}.
What is bound to `\texttt{\$pat}' does not matter in the pattern of the \texttt{match-all} expression, too.
We can use a \textit{variable-pattern} in the body of a pattern-function.
We cannot use it in the pattern of a match-clause.
In the following sample, `\texttt{pat1}' and `\texttt{pat2}' in the body of \texttt{twin} are variable-patterns.
They must be the arguments of the pattern-function.

{\footnotesize
\begin{verbatim}
> (define $twin
    (pattern-function [$pat1 $pat2]
      <cons (& $pat pat1) <cons ,pat pat2>>))
> (match-all {1 2 1 3} (multiset integer) [<cons $m (twin $n _)> [m n]]))
{[2 1] [2 1] [3 1] [3 1]}
\end{verbatim}
}

My pattern-matching system restricts use of patterns in match-clauses and bodies of pattern-functions.
This restriction enables us to reuse our own patterns in a simple way.
If we treat a pattern as a first class object as first class patterns~\cite{tullsen2000first},
it is difficult to modularize patterns that contain pattern variables.

A pattern-function can take only patterns.
If we would like to write a pattern that takes parameters, we write a function that obtains the objects as parameters and returns a pattern-function.

Figure~\ref{fig:mahjong} is a demonstration program that determines whether mahjong hands are finished or not.
It is a very hard task if we write it without my pattern-matching system.
I explain the definitions for each matcher in section~\ref{matcher}.

\begin{figure}
  \begin{center}

  {\scriptsize
\begin{verbatim}
;;
;; Matcher definitions
;;
(define $suit
  (algebraic-data-matcher
    {<wan> <pin> <sou>}))

(define $honor
  (algebraic-data-matcher
    {<ton> <nan> <sha> <pe> <haku> <hatsu> <chun>}))

(define $tile
  (algebraic-data-matcher
    {<num suit integer> <hnr honor>}))

;;
;; Pattern modularization
;;
(define $twin
  (pattern-function [$pat1 $pat2]
    <cons (& $pat pat1)
     <cons ,pat
      pat2>>))

(define $shuntsu
  (pattern-function [$pat1 $pat2]
    <cons (& <num $s $n> pat1)
     <cons <num ,s ,(+ n 1)>
      <cons <num ,s ,(+ n 2)>
       pat2>>>))

(define $kohtsu
  (pattern-function [$pat1 $pat2]
    <cons (& $pat pat1)
     <cons ,pat
      <cons ,pat
       pat2>>>))

;;
;; A function that determines whether the hand is completed or not.
;;
(define $complete?
  (match-lambda (multiset tile)
    {[(twin $th_1
       (| (shuntsu $sh_1 (| (shuntsu $sh_2 (| (shuntsu $sh_3 (| (shuntsu $sh_4 <nil>)
                                                                (kohtsu $kh_1 <nil>)))
                                              (kohtsu $kh_1 (kohtsu $kh_2 <nil>))))
                            (kohtsu $kh_1 (kohtsu $kh_2 (kohtsu $kh_3 <nil>)))))
          (kohtsu $kh_1 (kohtsu $kh_2 (kohtsu $kh_3 (kohtsu $kh_4 <nil>)))))
       (twin $th_2 (twin $th_3 (twin $th_4 (twin $th_5 (twin $th_6 (twin $th_7 <nil>)))))))
      #t]
     [_ #f]}))
\end{verbatim}
  }

  \end{center}
  \caption{Pattern-matching against mahjong tiles}
  \label{fig:mahjong}
\end{figure}

\subsection{Formal Definition of Patterns}

This is the formal definition of the syntax of patterns.
Note that variable-patterns appear only in the body of pattern-functions.

\begin{grammar}
<pattern> ::= `_' \hfill (wildcard)
  \alt `$' <ident> \hfill (pattern-variable)
  \alt `,' <expr> \hfill (value-pattern)
  \alt `<' <ident> <pattern>* `>' \hfill (inductive-pattern)
  \alt `(' <pat-func-expr> <pattern>* `)' \hfill (pattern-application)
  \alt `(| ' <pattern>* `)' \hfill (or-pattern)
  \alt `(& ' <pattern>* `)' \hfill (and-pattern)
  \alt `^' <pattern> \hfill (not-pattern)
  \alt `[' <pattern>* `]' \hfill (tuple-pattern)

<pat-func-expr> ::= `(pattern-function' `[' <pat-var>* `]' <pattern'> `)'

<pattern'> ::= <ident> \hfill (variable-pattern)
  \alt `_' \hfill (wildcard)
  \alt `$' <ident> \hfill (pattern-variable)
  \alt `,' <expr> \hfill (value-pattern)
  \alt `<' <ident> <pattern'>* `>' \hfill (inductive-pattern)
  \alt `(' <pat-func-expr> <pattern'>* `)' \hfill (pattern-application)
  \alt `(|' <pattern'>* `)' \hfill (or-pattern)
  \alt `(&' <pattern'>* `)' \hfill (and-pattern)
  \alt `^' <pattern'> \hfill (not-pattern)
  \alt `[' <pattern'>* `]' \hfill (tuple-pattern)
\end{grammar}

\section{Matcher Definitions}\label{matcher}

We define matchers to specify how to pattern-match against each data type.
In this section, I explain how to define matchers.

\subsection{Overview}

Before explaining formal definition of matchers, I give an overview of the role of a matcher definition.

Briefly stated, the pattern-matching process can be regarded as a reduction process of stacks of \textit{matching-atoms}, which are tuples of a pattern, target and matcher.
In each step of the pattern-matching process, a matching-atom is popped out and multiple matching-atoms are pushed to the stack.
In a matcher definition, we define how a popped matching-atom is reduced to the matching-atoms that are pushed to the stack.

A function that obtains a popped matching-atom and returns the reduced matching-atoms is called a \textit{match-function}.
The following is a sample of input and output of the match-function.
`\texttt{M}' denotes a match-function.

{\footnotesize
\begin{verbatim}
M [<nil> {} (multiset integer)] = { {[[] [] []]} }

M [<cons $x $xs> {} (multiset integer)] = {}
  
M [<cons $x $xs> {1 2 3} (multiset integer)] =
  { {[$x 1 integer] [$xs {2 3} integer]}
    {[$x 2 integer] [$xs {1 3} integer]}
    {[$x 3 integer] [$xs {1 2} integer]} }
\end{verbatim}
}

\subsection{Formal Definition and Simple Examples}

I explain syntax to define matchers.

\begin{grammar}
<matcher-expr> ::= `(matcher' `\{' <primitive-pmc>* `\})'

<primitive-pmc> ::= `[' <primitive-pp> <next-matcher-expr> `\{' <primitive-dmc>* `\}]'

<primitive-dmc> ::= `[' <primitive-dp> <expr> `]'

<primitive-pp> ::= `\$'                \hfill (primitive-pattern-variable)
  \alt `,\$' <ident>                   \hfill (value-pattern-pattern)
  \alt `<' <ident> <primitive-pp>* `>' \hfill (primitive-inductive-pattern)
  
<primitive-dp> ::= `\_' \hfill (wildcard)
  \alt `\$' <ident> \hfill (primitive-data-variable)
  \alt `<' <Ident> <primitive-dp>* `>' \hfill (primitive-inductive-data)
  \alt `\{\}' \hfill (primitive-empty-collection)
  \alt `\{' <primitive-dp> `@' <primitive-dp> `\}' \hfill (primitive-cons-collection)
  \alt `\{' `@' <primitive-dp> <primitive-dp> `\}' \hfill (primitive-snoc-collection)
\end{grammar}

\textit{primitive-pmc} and \textit{primitive-dmc} are abbreviations of \textit{primitive-pattern-match-clause} and \textit{primitive-data-match-clause}, respectively.
\textit{primitive-pp} and \textit{primitive-dp} are abbreviations of \textit{primitive-pattern-pattern} and \textit{primitive-data-pattern}, respectively.
A primitive-pattern-pattern is a pattern that pattern-matches against a pattern.
A primitive-data-pattern is a pattern that pattern-matches against a target datum.
\textit{Ident} stands for an identifier that begins with an uppercase letter.
\textit{ident} stands for an identifier that begins with an lowercase letter.

Here is the first sample of a matcher definition.
With \texttt{unordered-pair}, we can pattern-match a pair of data ignoring the order of the elements of the pair.
For example, the datum \texttt{<Pair 2 5>} is pattern-matched with the pattern \texttt{<pair ,5 \$x>}.

{\footnotesize
\begin{verbatim}
> (define $unordered-pair
   (lambda [$a]
    (matcher {[<pair $ $> [a a] {[<Pair $x $y> {[x y] [y x]}]}]
              [$ [something] {[$tgt {tgt}]}]})))

> (match-all <Pair 2 5> (unordered-pair integer) [<pair ,5 $x> x])
{2}
> (match-all <Pair 2 5> (unordered-pair integer) [$p p])
{<Pair 2 5>}
\end{verbatim}
}

\texttt{unordered-pair} is defined as a function that gets a matcher and returns a matcher.
It is to specify how to pattern-match against the elements of the pair.

The \texttt{matcher} expression defines the way of pattern-matching for unordered-pairs.
First, the pattern is pattern-matched with each primitive-pattern-pattern.
The pattern \texttt{<pair ,5 \$x>} matches with the primitive-pattern-pattern \texttt{<pair \$ \$>}.
`\texttt{\$}' is called a \textit{primitive-pattern-variable}.
The first `\texttt{\$}' pattern-matches with `\texttt{,5}' and the second `\texttt{\$}' matches with `\texttt{\$x}'.
The patterns bound to `\texttt{\$}' are called \textit{next-patterns}.
We can create as many next-patterns as we want.
\texttt{[a a]} is a \textit{next-matcher-expression}.
A next-matcher-expression returns a tuple of matchers.
These are called \textit{next-matchers}.
In this case, `\texttt{a}' is bound to \texttt{integer}.
Therefore, both `\texttt{,5}' and `\texttt{\$x}' are pattern-matched as \texttt{integer}.
\texttt{[<Pair \$x \$y> \{[x y] [y x]\}]} is a primitive-data-match-clause.
\texttt{<Pair \$x \$y>} is pattern-matched with the target datum \texttt{<Pair 2 5>},
and `\texttt{\$x}' and `\texttt{\$y}' is matched with `\texttt{2}' and `\texttt{5}', respectively.
The pattern-matching of primitive-data-patterns is similar with the pattern-matching of the ordinary functional programming languages.
The primitive-data-match-clause returns \texttt{\{[2 5] [5 2]\}}.
The primitive-data-match-clause returns a collection of \textit{next-targets}.
This means the patterns `\texttt{,5}' and `\texttt{\$x}' are matched with the targets `\texttt{2}' and `\texttt{5}' or `\texttt{5}' and `\texttt{2}' using the \texttt{integer} matcher, respectively.

The pattern of the second \texttt{match-all} expression is a single pattern-variable.
In this case, the first primitive-pattern-pattern \texttt{<pair \$ \$>} does not match and the second primitive-pattern-pattern `\texttt{\$}' matches with the pattern and the second primitive-pattern-match-clause is used.
The next-matcher is \texttt{something}.
The next-pattern and the next-target do not change.
\texttt{something} is the built-in matcher of the pattern-matching system.
It can match only with a wildcard or a pattern variable.
it is used to bind a value to a pattern-variable.

Next, I introduce the matcher of integers.
`\texttt{eq?}' is a built-in function that determines equality of built-in data.
It returns `\texttt{\#t}' if two arguments are equal, otherwise it returns `\texttt{\#f}'.

{\footnotesize
\begin{verbatim}
(define $integer
 (matcher {[,$n [] {[$tgt (if (eq? tgt n) {[]} {})]}]
           [$ [something] {[$tgt {tgt}]}]}))
\end{verbatim}
}

In the definition of \texttt{integer}, there is an example of a value-pattern-pattern.
The primitive-pattern-pattern \texttt{,\$n} is a value-pattern-pattern.
The value bounded to the variable \texttt{n} can be referred in the body of the primitive-data-match-clause.
There are no next-patterns.
The next-matcher is an empty tuple.
If the pattern-matching succeeds, the next-target is a collection consists of an empty tuple.
Otherwise, the next-target is an empty collection.

We can define the \texttt{mod} matcher as follow.
\texttt{mod} is a function that takes a number and returns a matcher.
`\texttt{(mod m)}' is a matcher for the quotient ring modulo `\texttt{m}'.

{\footnotesize
\begin{verbatim}
(define $mod
 (lambda [$m]
  (matcher
    {[,$n [] {[$tgt (if (eq? (modulo tgt m) (modulo n m)) {[]} {})]}]
     [$ [something] {[$tgt {tgt}]}]
     })))
\end{verbatim}
}

We can define the \texttt{card} matcher that has appeared in the poker hands analyzer in figure~\ref{fig:poker} using \texttt{mod}.

{\footnotesize
\begin{verbatim}
(define $suit
 (matcher {[<spade> [] {[<Diamond> {[]}] [_ {}]}]
           [<heart> [] {[<Heart> {[]}] [_ {}]}]
           [<club> [] {[<Club> {[]}] [_ {}]}]
           [<diamond> [] {[<Diamond> {[]}] [_ {}]}]}))

(define $card
 (matcher {[<card $ $> [suit (mod 13)] {[<Card $s $n> {[s n]}]}]
           [$ [something] {[$tgt {tgt}]}]}))
\end{verbatim}
}

Matcher definitions for algebraic data types are verbose.
I have prepared a syntax sugar for that.

\begin{grammar}
<algebraic-data-matcher-expr> ::= `(algebraic-data-matcher {' <constructor-definition>* `})'

<constructor-definition> ::= `<' <ident> <matcher-expr>* `>'
\end{grammar}

We define the \texttt{suit} and \texttt{card} matcher using the \texttt{algebraic-data-matcher} expression in figure~\ref{fig:poker}.
We also define the \texttt{tile} matcher that has appeared in figure~\ref{fig:mahjong} using the \texttt{algebraic-data-matcher} expression.

\subsection{Matchers for Collection Data Types}

In this section, I explain matchers that handle collections.

My pattern-matching system handles collections as primitive and prepares primitive-pattern-patterns for them.
The \texttt{list} matcher is defined using them.

{\footnotesize
\begin{verbatim}
(define $list
 (lambda [$a]
  (matcher
   {[,$val []
     {[$tgt (match [val tgt] [(list a) (list a)]
              {[[<nil> <nil>] {[]}]
               [[<cons $x $xs> <cons ,x ,xs>] {[]}]
               [[_ _] {}]})]}]
    [<nil> [] {[{} {[]}] [_ {}]}]
    [<cons $ $> [a (list a)] {[{$x @$xs} {[x xs]}] [_ {}]}]
    [<join $ $> [(list a) (list a)]
     {[$tgt (letrec {[$unjoin'
                      (lambda [$xs $ys]
                       (match ys (list a)
                        {[<nil> {[xs {}]}]
                         [<cons $y $rs> {[xs ys] @(unjoin' {@xs y} rs)}]
                         }))]}
             (unjoin' {} tgt))]}]
    [$ [something] {[$tgt {tgt}]}]})))
\end{verbatim}
}

A Definition of the matcher \texttt{multiset} and \texttt{set} are given as follows.
We can define them simply using the \texttt{list} matcher.

{\footnotesize
\begin{verbatim}
(define $multiset
 (lambda [$a]
  (matcher
   {[,$val []
     {[$tgt (match [val tgt] [(list a) (multiset a)]
              {[[<nil> <nil>] {[]}]
               [[<cons $x $xs> <cons ,x ,xs>] {[]}]
               [[_ _] {}]})]}]
    [<nil> [] {[{} {[]}] [_ {}]}]
    [<cons $ $> [a (multiset a)]
     {[$tgt (match-all tgt (list a)
              [<join $hs <cons $x $ts>> [x {@hs @ts}]])]}]
    [$ [something] {[$tgt {tgt}]}]})))
\end{verbatim}
}

{\footnotesize
\begin{verbatim}
(define $set
  (lambda [$a]
    (matcher
      {[<nil> [] {[{} {[]}] [_ {}]}]
       [<cons $ $> [a (set a)]
        {[$tgt (match-all tgt (list a)
                 [<join _ <cons $x _>> [x tgt]])]}]
       [$ [something]
        {[$tgt {tgt}]}]
       })))
\end{verbatim}
}

Note the importance of value-pattern-patterns, we cannot realize non-linear pattern-matching without them.
Value-pattern-patterns is used to define how to handle values in patterns for each matcher.
Please also note that, we can use value-patterns not only to handle equality but also order.

\section{Mechanism of Pattern-Matching}\label{mechanism}

In this section, I explain the implementation of my pattern-matching system.

\subsection{Notions}

I introduce several notions to explain the mechanism of pattern-matching.
Here is a really brief explanation of each notion.
We will deepen the understanding of these notions, examining the examples in the following sections.

\begin{description}
\item[Matching-State]

  The pattern-matching process is reduction of a collection of \textit{matching-states}.
  Each matching-state has a stack of matching-trees and data to proceed pattern-matching.

\item[Matching-Tree]

  A \textit{matching-tree} has two kinds of forms, a matching-atom and a matching-node.
  
\item[Matching-Atom]
  
  A \textit{matching-atom} is a tuple of a pattern, a target, and a matcher.
  
\item[Matching-Node]

  A \textit{matching-node} has a stack of matching-trees as a matching-state.
  The structure of matching-node is similar with a matching state.

\end{description}

\subsection{Simple Non-Linear Patterns}

In this section, I explain how pattern-matching is executed for a simple non-linear pattern.
Let us examine what will happen when the system evaluates the following pattern-matching expression.

{\footnotesize
\begin{verbatim}
> (match-all {2 8 2} (multiset integer)
   [<cons $m <cons ,m _>> m])
{2 2}
\end{verbatim}
}

At first, the initial matching-state is generated.
It is as follow.
The data constructor \texttt{MState} takes three arguments, a stack of matching-trees, an environment, and a result in the middle of the pattern-matching.
The `\texttt{env}' is the environment when the evaluation process enters the \texttt{match-all} expression.
The stack of matching-tree contains a single matching-atom whose pattern, target and matcher are same with the arguments of the \texttt{match-all} expression.

{\footnotesize
\begin{verbatim}
MState {[<cons $m <cons ,m _>> {2 8 2} (multiset integer)]} env {}
\end{verbatim}
}

The stack of the matching-tree contains only one matching-atom.
This matching-atom is reduced with the matcher `\texttt{(multiset integer)}' as specified in the matching-atom.
The matching-states increases to 3 with this reduction as follow.

{\footnotesize
\begin{verbatim}
MState {[$m 2 integer] [<cons ,m _> {8 2} (multiset integer)]} env {}

MState {[$m 8 integer] [<cons ,m _> {2 2} (multiset integer)]} env {}

MState {[$m 2 integer] [<cons ,m _> {2 8} (multiset integer)]} env {}
\end{verbatim}
}

We focus on the first matching-state, for now.
This matching-state is reduced as follow in the next reduction step.
The matcher of the matching-atom of the top of the stack is changed to \texttt{something} from \texttt{integer}.
\texttt{something} is the built-in matcher of the pattern-matching system.
It can match only with a wildcard or a pattern variable.
it is used to bind a value to a pattern-variable.

{\footnotesize
\begin{verbatim}
MState {[$m 2 something] [<cons ,m _> {8 2} (multiset integer)]} env {}
\end{verbatim}
}

This matching-state is reduced as follow in the next reduction step.
The matching atom of the top of the stack is popped out, and a new binding \texttt{[m 2]} is appended to the result of the middle of pattern-matching.
\texttt{something} can only append a new binding to the result of pattern-matching.

{\footnotesize
\begin{verbatim}
MState {[<cons ,m _> {8 2} (multiset integer)]} env {[m 2]}
\end{verbatim}
}

This matching-state is reduced as follow in the next reduction step.
The matching-states increases to 2 with this reduction.

{\footnotesize
\begin{verbatim}
MState {[,m 8 integer] [_ {2} (multiset integer)]} env {[m 2]}

MState {[,m 2 integer] [_ {8} (multiset integer)]} env {[m 2]}
\end{verbatim}
}

In the above matching-states, `\texttt{,m}' is pattern-matched with `\texttt{8}' and `\texttt{2}' respectively as \texttt{integer}.
When we pattern-match with a value pattern, the result of the middle of pattern-matching is used to evaluate it.
Therefore, in this case, `\texttt{m}' is evaluated to `\texttt{2}'.
The first matching-state fails to pattern-match and vanishes.
The second matching-state succeeds in pattern-matching and be reduced as follow in the next reduction step.

{\footnotesize
\begin{verbatim}
MState {[_ {8} (multiset integer)]} env {[m 2]}
\end{verbatim}
}

This matching-state is reduced as follow in the next reduction step.
The pattern is a wildcard and matches with any object.
No new binding is appended to the result of pattern-matching.

{\footnotesize
\begin{verbatim}
MState {} env {[m 2]}
\end{verbatim}
}

When the matching-tree stack is empty, the reduction finishes and the matching-state succeeds in patter-matching.
This result of pattern-matching \texttt{{[m 2]}} is added to the final result.

\subsection{Or-Patterns, And-Patterns, Not-Patterns and Tuple-Patterns}\label{or-and-not-patterns}

Or-patterns, and-patterns and not-patterns are specially handled.
In this section, we explain them.

\subsubsection{Or-Patterns}

Let us examine what will happen when the system evaluates the following pattern-matching expression.

{\footnotesize
\begin{verbatim}
> (match-all {1 1 2} (list integer)
   [<cons $m (| <nil> <cons ,m _>)> m])
{1}
\end{verbatim}
}

The system reaches the following matching-state.

{\footnotesize
\begin{verbatim}
MState {[(| <nil> <cons ,m _>) {1 2} (list integer)]} env {[m 1]}
\end{verbatim}
}

This matching-state is reduced as follow in the next reduction step.

{\footnotesize
\begin{verbatim}
MState {[<nil>       {1 2} (list integer)]} env {[m 1]}

MState {[<cons ,m _> {1 2} (list integer)]} env {[m 1]}
\end{verbatim}
}

\subsubsection{And-Patterns}

Let us examine what will happen when the system evaluates the following pattern-matching expression.

{\footnotesize
\begin{verbatim}
> (match-all {1 2 3} (list integer)
   [<cons $n (& <cons _ _> $rs)> [n rs]])
{[1 {2 3}]}
\end{verbatim}
}

The system reaches the following matching-state.

{\footnotesize
\begin{verbatim}
MState {[(& <cons _ _> $rs) {2 3} (list integer)]} env {[n 1]}
\end{verbatim}
}

This matching-state is reduced as follow in the next reduction step.

{\footnotesize
\begin{verbatim}
MState {[<cons _ _> {2 3} (list integer)]
        [$rs        {2 3} (list integer)]}
       env {[n 1]}
\end{verbatim}
}

\subsubsection{Not-Patterns}

Let us examine what will happen when the system evaluates the following pattern-matching expression.

{\footnotesize
\begin{verbatim}
> (match-all {2 8 2} (multiset integer)
   [<cons $m (& ^<cons ,m _> $rs)> [m rs]])
{[8 {2 2}]}
\end{verbatim}
}

The system reaches the following matching-state.

{\footnotesize
\begin{verbatim}
MState {[^<cons ,m _> {2 2} (multiset integer)]
        [$rs          {2 2} (multiset integer)]}
       env {[m 8]}
\end{verbatim}
}

When the system reaches the matching-state whose top matching-atom is a not-pattern,
the system generates a new matching-state that contains only the matching-atom with the not-pattern as follow.
All information of the matching-state and the matching-nodes except about the rest of matching-tree stack are retained.

{\footnotesize
\begin{verbatim}
MState {[<cons ,m _> {2 2} (multiset integer)]} env {[m 8]}
\end{verbatim}
}

The system proceeds the pattern-matching on the new generated matching-state,
and if it fails pattern-matching the system pops out the matching-atom of the not-pattern from the original matching-state as follow and proceeds the pattern-matching process.
Otherwise the matching-state fails pattern-matching.

In this case, the above matching-state fails pattern-matching.
Therefore, the matching-state with the not-pattern is reduced as follow.

{\footnotesize
\begin{verbatim}
MState {[$rs {2 2} (multiset integer)]} env {[m 8]}
\end{verbatim}
}

\subsubsection{Tuple-Patterns}

Let us examine what will happen when the system evaluates the following pattern-matching expression.

{\footnotesize
\begin{verbatim}
> (match-all [3 3] [integer integer]
   [[$n ,n] n])
{3}
\end{verbatim}
}

The initial matching-state is as follow.

{\footnotesize
\begin{verbatim}
MState {[[$n ,n] [3 3] [integer integer]]} env {}
\end{verbatim}
}

This matching-state is reduced as follow in the next reduction step.

{\footnotesize
\begin{verbatim}
MState {[$n 3 integer]
        [,n 3 integer]}
       env {}
\end{verbatim}
}

\subsection{Application of Pattern-Functions}\label{mechanism-of-pattern-functions}

In this section, I explain how the system deals with modularization of patterns.
Let us examine what will happen when the system evaluates the following pattern-matching expression.

{\footnotesize
\begin{verbatim}
> (define $twin
    (pattern-function [$pat1 $pat2]
      <cons (& $pat pat1)
       <cons ,pat
        pat2>>))
> (match-all {1 2 1 3} (multiset integer)
      [<cons $m (twin $n _)> [m n]])
{[2 1] [3 1] [2 1] [3 1]}
\end{verbatim}
}

The system reaches the following matching-state.

{\footnotesize
\begin{verbatim}
MState {[(twin $n _) {1 1 3} (multiset integer)]} env {[m 2]}
\end{verbatim}
}

This matching-state is reduced as follow in the next reduction step.
A matching-node has extra information, a \textit{pattern-environment}.
In this case, the pattern-environment is \texttt{\{[pat1 \$n] [pat2 _]\}}.

{\footnotesize
\begin{verbatim}
MState {(MNode {[<cons (& $pat pat1) <cons ,pat pat2>>
                 {1 1 3} (multiset integer)]}
               env1 {} {[pat1 $n] [pat2 _]})}
       env {[m 2]}
\end{verbatim}
}

The system reaches the following matching-state.
When the top of the matching-tree stack of the matching-state is a matching-node, the system pops the matching-atom of the top of the matching-tree stack of the matching-node.
If the top of the matching-tree stack of the matching-node is a matching-node again, the system pops out the matching-atom from the top of the matching-tree stack of that matching-node.

{\footnotesize
\begin{verbatim}
MState {(MNode {[pat1 1 integer]
                [<cons ,pat pat2> {1 3} (multiset integer)]}
               env1 {[pat 1]} {[pat1 $n] [pat2 _]})}
       env {[m 2]}
\end{verbatim}
}

This matching-state is reduced as follow in the next reduction step.
\texttt{pat1} is called a \textit{variable-pattern}.
It can appear only in the body of pattern-functions.
When the matching-atom whose pattern is a variable-pattern is popped out,
the system gets what pattern is bound to the variable-pattern from the pattern-environment,
and push a new matching-atom to the matching-tree stack of the one level upper matching-node or matching-state.

{\footnotesize
\begin{verbatim}
MState {[$n 1 integer]
        (MNode {[<cons ,pat pat2> {1 3} (multiset integer)]}
               env1 {[pat 1]} {[pat1 $n] [pat2 _]})}
       env {[m 2]}
\end{verbatim}
}

The arguments of a pattern-function are handled in special way as above.
This is the reason why the pattern-function can take only patterns.
A pattern must be bound to a variable-pattern.

\subsection{Pattern-Matching with Infinite Results}

In this section, I explain how the system executes pattern-matching that has infinite results.
Let us examine what will happen when the system evaluate the following pattern-matching expression.

{\footnotesize
\begin{verbatim}
> (take 10 (match-all nats (set integer) [<cons $m <cons $n _>> [m n]]))
{[1 1] [1 2] [2 1] [1 3] [2 2] [3 1] [1 4] [2 3] [3 2] [4 1]}
\end{verbatim}
}

\begin{figure}
  \begin{center}
    \includegraphics[width=8cm]{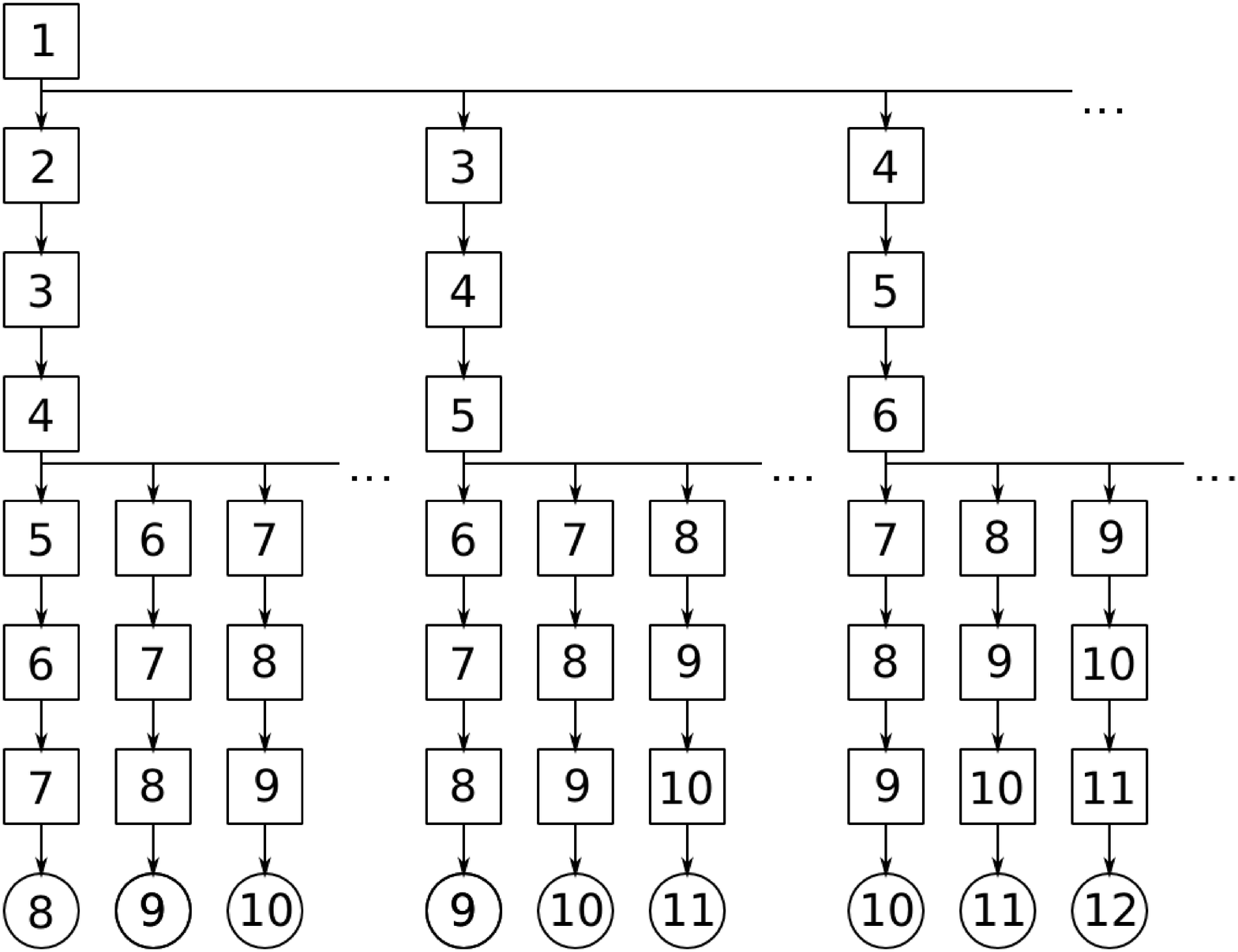}
  \end{center}
  \caption{Reduction tree of matching-states}
  \label{fig:reduction}
\end{figure}

Figure~\ref{fig:reduction} is the reduction tree of matching-states when we execute the above pattern-matching expression.
Rectangles stand for matching-states.
The rectangle at the upper left is the initial matching-state.
Circles stand for final matching-states that succeed pattern-matching.

The width of a reduction tree of matching-states can be infinite because there are cases that a matching-state is reduced to infinite matching-states.
The depth of a reduction tree also can be infinite if we use a recursive pattern-function in a pattern.
We need to think on the order of reduction to examine all nodes of a reduction tree.
The numbers on rectangles and circles denote the order of reduction.
If we see a reduction tree obliquely, it can be regarded as a binary tree.
Therefore, we can trace all nodes of reduction trees if we do breadth-first search on the tree, though it will use a lot of memory.

\section{Related Work}

In this section, I introduce existing studies in the field of pattern-matching.


Miranda laws~\cite{thompson1990lawful,turner1985miranda} and Wadler's views~\cite{wadler1987views} are famous work.
These proposals provide the way to decompose data that have multiple representations, by declaring transformation between each representations.
For example, we can intuitively handle complex numbers that have cartesian and polar representation.

{\footnotesize
\begin{verbatim}
complex ::= Pole real real

view complex ::= Cart real reel
  in  (Pole r t) = Cart (r * (cos t)) (r * (sin t))
  out (Cart x y) = Pole (sqrt (x * x + y * y)) (atan2 x y)

add  (Cart x y) (Cart x’ y’) = Cart (x + x’) (y + y’)
mult (Pole r t) (Pole r’ t’) = Pole (r * r’) (t + t’)
\end{verbatim}
}

Data are automatically transformed in the matching process.
We define the way of transformation for each combination of data constructors.
However, the pattern-matching systems of these proposals treat neither multiple results of pattern-matching nor non-linear patterns.
These studies demand a canonical form for each representation.

Active patterns~\cite{erwig1996active} provide a way to decompose unfree data.
We define a \textit{match function} for each pattern constructor to decompose unfree data.
In the following sample code, \texttt{Add'} is a match function.
With the match function \texttt{Add'}, we can extract an element ignoring the order of elements from the target that is constructed with the \texttt{Add} constructor.

{\footnotesize
\begin{verbatim}
pat Add' (x,_) =
  Add (y,s) => if x == y then (y,s)
                        else let Add' (x,t) = s
                              in Add (x, Add (y, t)) end

fun member x (Add' (x,s)) = true
  | member x s            = false
\end{verbatim}
}

The \texttt{Add'} match function enables us to handle multisets directly.
The demonstrations of active patterns for pattern-matching against graphs are also proposed.~\cite{erwig1997functional}

The weakness of active patterns is that it does not support backtracking in the pattern-matching process.
The value bound to pattern variables must be fixed from the left side of a pattern,
though many forms should be tried for pattern-matching with unfree data types.
For example, we cannot write the pattern for a collection with multiple same elements.
It also does not support pattern-matching with multiple results.
We cannot write poker-hands pattern-matching in active patterns as figure~\ref{fig:poker}.

First class patterns~\cite{tullsen2000first} propose a sophisticated system that treats patterns as first class objects.
The essence of this study is a \textit{pattern-function} that defines how to decompose data with each data constructor.
In the following sample code, `\texttt{cons\#}' is a pattern-function.
The pattern function `\texttt{cons\#}' helps to decompose a list in the join representation.

{\footnotesize
\begin{verbatim}
data List a = Nil | Unit a | Join (List a) (List a)

cons x xs = Join (Unit x) xs

cons# Nil = Nothing
cons# (Unit a) = Just (a,Nil)
cons# (Join xs ys) = case cons# xs of
                       Just (x,xs') -> Just (x, Join xs' ys)
                       Nothing      -> cons# ys
\end{verbatim}
}

First class patterns can deal with pattern-matching that generates multiple results.
To generate multiple results, a pattern-function returns a list, not a datum of the type \texttt{Maybe}.
However pattern-matching with this proposal also has a weak point.
First class patterns do not support non-linear pattern-matching, though non-linear patterns are necessary to express meaningful patterns for unfree data types.

My proposal can be seen as extension of these researches.

Functional logic programming~\cite{antoy2010functional} is different approach.
It uses the unification mechanism of logic programming to decompose data.
We can pattern-match against unfree data types because the unification system has a backtracking mechanism.
However, the pattern-matching expressions of functional logic programming are not intuitive compared with the other proposals.
For example, the following sample defines a function to determine whether the argument cards satisfy four-of-kinds.~\cite{antoy2010programming}

{\footnotesize
\begin{verbatim}
four (x++[y]++z) | map rank (x++z) =:= [r,r,r,r]
                 = r
                 where r free
\end{verbatim}
}

One of the reasons the above program is not intuitive is it treats the collection of the cards as a list, not as a multiset directly.
We cannot modularize the way of pattern-matching for each data type.
We have very strong expressive power in a guard in functional logic programming.
However, the expressive of power of pattern-matching is not strong.

\section{Conclusion}

The combination of \textbf{all of the following features} realizes intuitive powerful pattern-matching.

\begin{description}
  \item \textbf{Modularization of the way of pattern-matching}

    We can define the way of pattern-matching for each data type.
    For example, we can define how to pattern-match against lists, multiset, and sets respectively.
    One of the characteristics of my proposal is that pattern-matching methods are specified by matchers for each data type not for each pattern-constructor.
    It enables us to reuse pattern-constructors and pattern-functions for similar data types.
    For example, we can use the same pattern-constructors and pattern-functions, such as \texttt{nil}, \texttt{cons} and \texttt{twin} for lists and multisets.
    This is very useful because unfree data are often pattern-matched using different matchers in different places of the program.

  \item \textbf{Multiple pattern-matching results}

    We can handle pattern-matching that has multiple results with backtracking.
    This feature is necessary to pattern-matching against data types whose data have no standard form.

  \item \textbf{Non-linear patterns}

    We can handle multiple occurrences of same variables in a pattern.
    Non-linear patterns are represented with value-patterns that match if the target is equal with the content of the pattern.
    Non-linear pattern-matching is realized with a rule that pattern-matching is executed from the left side of the pattern.
\end{description}

Furthermore, we realized pattern modularization with lexical scoping.
Lexical scoping in patterns became difficult because of the above features.

\begin{description}
  \item \textbf{Lexical scoping of patterns}

    We modularize patterns with pattern-functions, functions that receive patterns and return a pattern.
    Since a pattern-function has lexical scoping, bindings for pattern-variables in the argument patterns and the body of pattern-functions don't conflict.
    Useful patterns can be reused in many places in a program without worry of name conflicts.
    The tree-shaped matching-tree stack mechanism realizes lexical scoping in patterns.
\end{description}


Finally, understanding of human's intuition and finding direct representation of it are very important in many fields of computer science especially for automated reasoning and automated programming.
I hope my work will make breakthroughs in these fields.

\section*{Acknowledgement.}
First, I would like to thank Kentaro Honda. He is the first user of Egison and always encourages me.
I would like to thank Masami Hagiya and Yoichi Hirai for their great support and advice.
I would like to thank Ryo Tanaka, Takahisa Watanabe and Takuya Kuwahara for their help to implement the interpreter.
I would like to thank Yoshihiko Kakutani, Ibuki Kawamata, Takahiro Kubota, Takasuke Nakamura, Tomoya Chiba, Shigekazu Takei, Masato Hagiwara, Takashi Umeda and Michal J. Gajda for their continual feedback.
I would like to thank Yasunori Harada, Ikuo Takeuchi and people in IPA for their support for the development of Egison and help to hold the first workshop in Tokyo.
I would like to thank Yi Dai for his great feedback on my paper.
Finally, I would like to thank colleagues in Rakuten Institute of Technology for their support and great advice.

\bibliographystyle{abbrvnat}
\bibliography{egison}

\end{document}